\documentclass[10pt,twocolumn]{IEEEtran}
\usepackage{amsmath}
\usepackage{amsfonts}
\usepackage{epsfig}
\usepackage{amssymb}
\usepackage{cite}
\usepackage{hhline}
\usepackage{multirow}
\usepackage{framed}
\usepackage{xcolor}
\usepackage{lipsum}
\usepackage{tabularx}
\usepackage{soul}
 \usepackage{url}

\usepackage{color}
\usepackage{graphicx}
\usepackage{subfigure}

\makeatletter
\def\hlinewd#1{%
\noalign{\ifnum0=`}\fi\hrule \@height #1 %
\futurelet\reserved@a\@xhline}
\makeatother

\begin{document}

\title{Green Sensing and Access: Energy-Throughput Tradeoffs in Cognitive Networking}

\author{Hossein Shokri-Ghadikolaei, Ioannis Glaropoulos, Viktoria Fodor, Carlo Fischione, and Antony Ephremides %
\thanks{H. Shokri-Ghadikolaei, C. Fischione, and V. Fodor are with the KTH Royal Institute of Technology, Stockholm,
Sweden (email: \{hshokri, carlofi, vjfodor\}@kth.se).}
\thanks{I. Glaropoulos is with the KTH Royal Institute of Technology and Swedish Institute of Computer Science in Stockholm,
Sweden (email: ioannisg@sics.se).}
\thanks{A. Ephremides is with the University of Maryland, College Park (email: etony@umd.edu)}
\thanks{This work was supported by the FP7 EU project Hydrobionets.}
}

\maketitle

\begin{abstract}
Limited spectrum resources and dramatic growth of high data rate applications have motivated opportunistic spectrum access utilizing the promising concept of cognitive networks. Although this concept has emerged primarily to enhance spectrum utilization and to allow the coexistence of heterogeneous network technologies, the importance of energy consumption imposes additional challenges, because energy consumption and communication performance can be at odds. In this paper, the approaches for energy efficient spectrum sensing and spectrum handoff, fundamental building blocks of cognitive networks is investigated. The tradeoff between energy consumption and throughput, under local as well as under cooperative sensing are characterized, and what further aspects need to be investigated to achieve energy efficient cognitive operation under various application requirements are discussed.
\end{abstract}

\section{Introduction}\label{Intro}
The popularity of devices such as smart phones, tablets, and laptops all wirelessly connected to the Internet, as well as the recent development of the Internet of Things paradigm introduces an ever-growing demand for spectrum along with the need for heterogeneous networking technologies that suit various networked applications. Cognitive networks and opportunistic spectrum access in licensed frequency bands may become a key technology to address those demands by increasing the spectral efficiency, providing enough resources to realize massive machine-type communication for billions interconnected devices forecasted for 2020\footnote{See \url{http://machinaresearch.com/forecasts/​}.}, and facilitating coexistence among diverse networks and integration into future cellular network.

There has been a substantial effort on the design of cognitive networks with focus on throughput maximization. At the same time, the importance of reduced energy consumption, both due to the operation costs and for supporting battery operated devices, impose new challenges. As reducing energy consumption may reduce communication performance, energy consumption optimization has to be considered with the target application quality requirements in mind. One of the main existing efforts to address this challenge is GreenTouch, a consortium of  academia, vendors, and operators, launched in 2010, with the mission of decreasing the per bit energy consumption to one-thousandth of that in 2010, by 2015\footnote{Detailed information is available on http://www.greentouch.org.}.

The investigation of energy efficient cognitive radio technology as a mean to increase the spectral efficiency of future wireless networks requires the understanding of the energy cost imposed by the functionalities related to the cognitive operation. Compared to traditional wireless networks, opportunistic spectrum access in a cognitive network requires appropriate spectrum sensing and spectrum handoff mechanisms, which may be a substantial source of energy consumption in a network with a large number of cognitive device. In general, more accurate sensing and handoff control demands higher energy consumptions, which can be justifiable if it leads to significant gain in the spectrum utilization, introducing a tradeoff between energy consumption and throughput. Our objective is to characterize this tradeoff and evaluate what parameters need to be considered to optimize the cognitive network operation in different networking environments. Based on the discussion of existing proposals, we evaluate the additional parameters such as cooperative sensing incentives that should be considered to allow energy efficient operation in large networks, where users may have different transmission needs and possibly conflicting interests.

\section{Fundamentals}\label{Fundamentals}

\subsection{Cognitive Networks for Opportunistic Spectrum Access}
Under opportunistic spectrum sharing, two or more networks share some part of the spectrum. The primary network, with several primary users (PUs), owns the spectrum. The secondary network(s), of secondary users (SUs), can access the spectrum if no significant degradation on the primary communication is caused, in terms of interference level, throughput, or delay. As secondary communication needs to take the state of the primary network into account, cognitive network operation is necessary.

\begin{figure}[t]
	\centering
		\includegraphics[width=8.5cm]{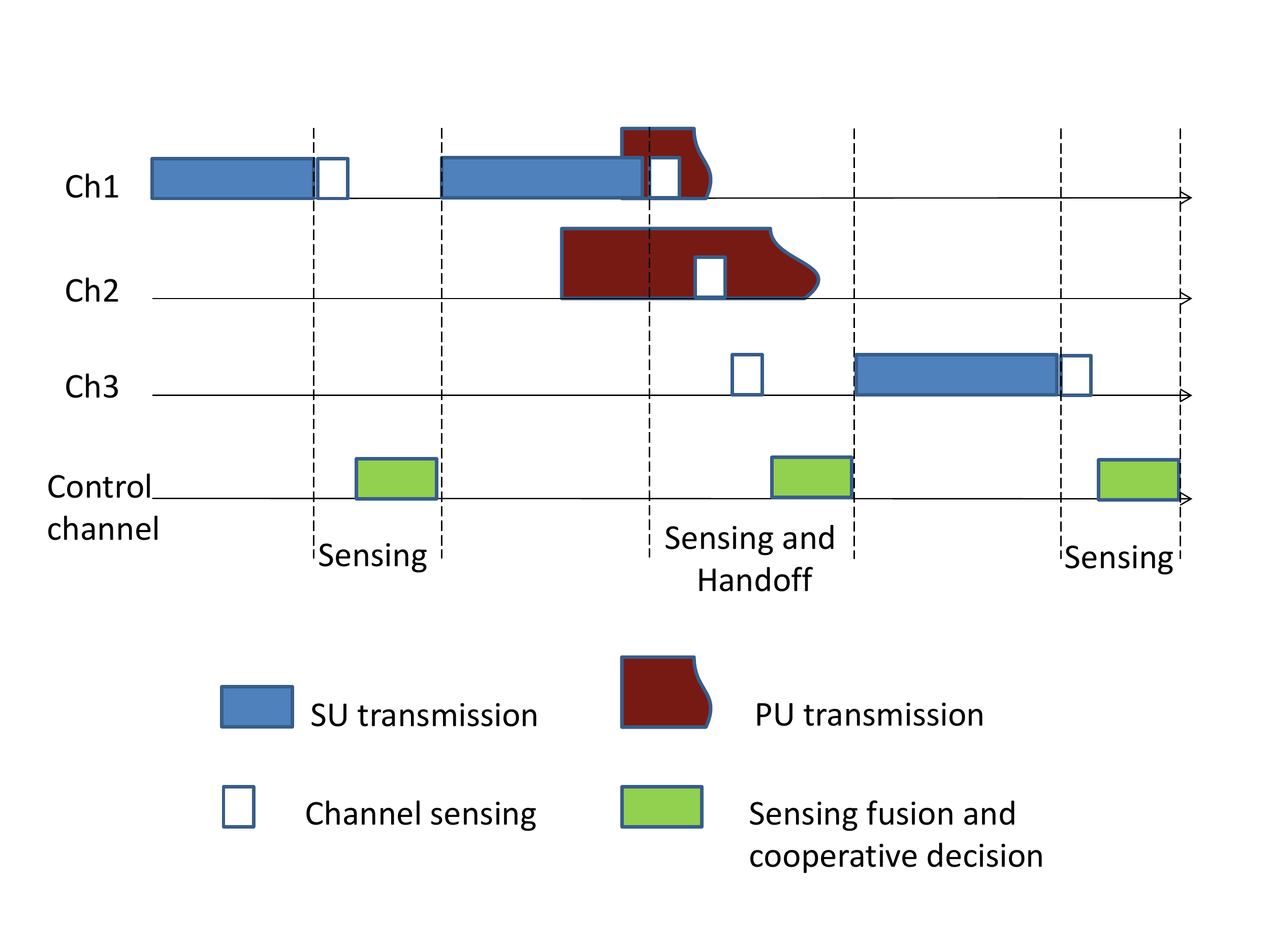}
	\caption{The SU interrupts transmission to evaluate the availability of the  channel using cooperative sensing. If the channel is busy, it starts a spectrum handoff procedure to find an idle channel. Spectrum sensing and handoff consume both energy and time.}
	\label{fig:crn-sensing}
\end{figure}

To find the opportunities of spectrum access, the cognitive secondary network learns the wireless environment and adapts to it. The learning is often based on sensing the spectrum, while the adaptation includes the tuning of  various parameters of the protocols.
As shown in Fig.~\ref{fig:crn-sensing}, to find the transmission opportunities appropriately and to protect the PUs from harmful interference, the SUs need to sense the channels regularly using local or cooperative sensing, and to start a spectrum handoff procedure, if the current channel is busy.

\subsection{Spectrum Sensing}
The most important parameters affecting  the performance of spectrum sensing are the time available to sense the transmission channels and the strength of the primary signals. A-priori information on the primary technology may determine which spectrum sensing method should be applied, from energy sensing to feature based detection schemes. Under all schemes, however, noise and channel impairments such as shadowing and fading lead to decision errors, quantified in terms of false alarm and misdetection probabilities. A false alarm occurs when a free channel is mistakenly sensed busy, while a misdetection happens whenever an occupied channel is sensed free. To improve the sensing performance, cooperative sensing may be introduced, where a group of SUs together decide about the availability of the channel, increasing the robustness of the spectrum sensing by utilizing the spatial diversity of the individual links.

\subsection{Multichannel Spectrum Sensing}
As typically there are more than one primary channels available for secondary access, spectrum sensing methods are generally classified into wideband and narrowband sensing.
Under wideband spectrum sensing, an SU senses multiple channels simultaneously. Although this may allow short sensing duration, it requires complex hardware implementation. Under narrowband sensing, only one channel can be sensed at a time, which allows simple sensing hardware and decision mechanisms, and therefore this is the preferred solution for most of the proposed cognitive systems. In this case, sensing time and sensing energy consumption may increase linearly with the number of sensed channels.





\subsection{Spectrum Handoff Strategy}
The spectrum handoff strategy answers the questions: when should an SU vacate the current channel? should the SU wait on this channel or start finding an available one? which channels should be sensed and in what order?

The strategies can be categorized as reactive, proactive, or, as a combination of these, hybrid.  
Under reactive spectrum handoff, the SU recognizes that a PU started to use the channel, and therefore it should vacate the channel. The SU then initiates searching among the channels to find transmission opportunities and pursue its unfinished transmission.
Although a larger delay becomes inevitable, reactive spectrum handoff builds on up-to-date channel status estimation.
Proactive schemes, on the other hand, exploit the long term traffic statistics of the channels to establish a proper policy for future spectrum handoffs. 
To allow detailed channel occupancy statistics, these schemes may require two radios, one for transmission and one for continuously scanning the channels.
Hybrid strategies combine the advantages of the two basic schemes, that is, prepare the sensing order of the channels in advance based on available statistics, but performing reactive channel sensing at handoff triggers to find an idle channel.

\subsection{Energy Efficiency}
Energy efficiency is generally defined as the number of information bits transmitted per unit of energy, measured in \emph{bit-per-Joule}. Alternatively, it is reflected by the energy cost, that is, the energy required to transmit a unit of information, measured in~\emph{Joule-per-bit}.
The energy consumed in the secondary network consists of consumption for i) data transmission and reception, ii)  spectrum sensing, and iii) the communication protocol to support the cognitive operation, including for example information exchange for cooperative spectrum sensing and for coordinating secondary transmissions. Finally, minor components are the circuit powers and the power consumed for tuning to a target channel~\cite{PeiJSAC2011,WangVT12}.

By the Shannon's capacity formula, it is known that in a dedicated spectrum, linearly increased transmission power leads to a logarithmic increase of the achievable transmission rate, and consequently the energy efficiency, as the ratio of the rate and the invested energy has an optimum value. 
The tradeoff between energy consumption and throughput becomes more complex in cognitive networks, as sensing consumes  energy, valuable time when the primary channel is idle, and also communication resources for cooperative sensing.

In Table~\ref{tab: comparision}, we summarize the solutions proposed to improve the energy efficiency of sensing and channel access both under local and under cooperative sensing. In the next sections, we discuss in detail the involved parameters, the effect of their optimization, and the implementation challenges. The presented results are based on a variety of primary technologies (DTV, LTE, IEEE~802.11, etc.), and therefore we discuss trends instead of quantitative results.


\begin{table*}[!t]
  \centering
\caption{Main design parameters to achieve energy efficient spectrum sensing and handoff strategies}
  \label{tab: comparision}
  { 
\renewcommand{\tabcolsep}{2pt}
  \renewcommand{\arraystretch}{2}
  \begin{tabular}{|p{0.15\textwidth}|p{0.37\textwidth}|p{0.37\textwidth}|p{0.07\textwidth}|}
  \hline
  \bf{Design parameters} & \bf{Applicability} & \bf{Characteristics} & \bf{Reference}  \\
  \hline
  Channel sensing time  &  Necessary in all networking scenarios; gain under loose PU interference constraints & Closed form based optimization, or quasi-concave optimisation under non-backlogged traffic   &\cite{ShokriIETC13,Fanous2014Access}   \\ \hline
  Waiting or handoff & Necessary to avoid unnecessary handoffs in multi-channel systems; gains under loose SU delay constraints and heterogeneous channel occupancy           & Convex optimization under homogeneous primary channels; suboptimal algorithms with polynomial complexity for heterogeneous channels; building on the channel occupancy statistics                  &\cite{WangVT12,zhang2013what} \\  \hline
  Sensing order   & Gains under heterogeneous channel occupancy in multi-channel systems  & Greedy or polynomial suboptimal algorithms; need learning of the channel occupancy statistics &\cite{PeiJSAC2011,fanous2014reliable} \\ \hline
  Handoff maximization & Significant gain for non-saturated SUs & Closed form based optimization &\cite{ShokriIETC13} \\ \hline
  Combined sensing and channel access & Important for uncoordinated SUs; efficient if SU load is not too high & Local optimization based on Markovian system model, suboptimal iterative algorithms & \cite{Shokri2015Analysis,song2012prospect} \\  \hline
  Cooperation with the PU & Important under spectrum shortage; significant gain if PU delay requirements are not strict & Local; convex optimization  &\cite{Costa2014Energy} \\
  \hline
  Cooperative sensing, resource allocation & Necessary for cooperative sensing; significant gain under known SU density  & Local numerical optimization based on analytic models  & \cite{saud2013,glaropoulos2014} \\ \hline
  Cooperative sensing, user selection & Necessary in cooperative sensing, significant gain under diverse and correlated local sensing performance  &  Efficient suboptimal greedy algorithms; possible distributed realizations; integrated with with the correlation estimation  & \cite{cacciapuoti2012,salim2013} \\  \hline
  Sensing report forwarding & Important if reporting costs are significant; improved \hspace{0.5cm} efficiency if channel occupancy statistics are available  & Efficient greedy algorithms if channel occupancy statistics are available, otherwise suboptimal iterative node selection algorithms and local learning &\cite{maleki2011} \\   \hline
  Decision combining & Optimal combining rule leads to significant gain if \hspace{1cm} reporting links are unreliable & Centralized, numerical analysis based optimization & \cite{chaudhari12} \\ \hline
  \end{tabular}
}
\end{table*}

\section{Local Spectrum Sensing and Handoff Strategies}\label{LocalSensing}
Local spectrum sensing can provide adequate sensing performance if the primary signals are sufficiently strong. 
In the following, we discuss how key design decisions, as per channel sensing time, number and order of handoffs, and the coordination sensing and channel access affect the energy efficiency of the cognitive network. The results we discuss typically consider energy detection based sensing, albeit more advanced sensing methodologies present similar tradeoffs.


\subsection{Channel Sensing Time}\label{SensingTimeSebSection}
Sensing time is the most basic parameter that affects energy efficiency. Increasing the time spent to sense a single channel improves the performance of the spectrum sensing at the expense of an increased energy consumption and a possible decreased transmission time of an SU. In multi-channel systems, accurate sensing with long sensing times may still be beneficial, as it can avoid unnecessary handoffs, leading to a reduction of the energy consumption of the overall sensing process as well as to an increase of the time available for transmission.

\begin{figure}[t]
	\centering
	\subfigure[]{
		\includegraphics[width=8cm]{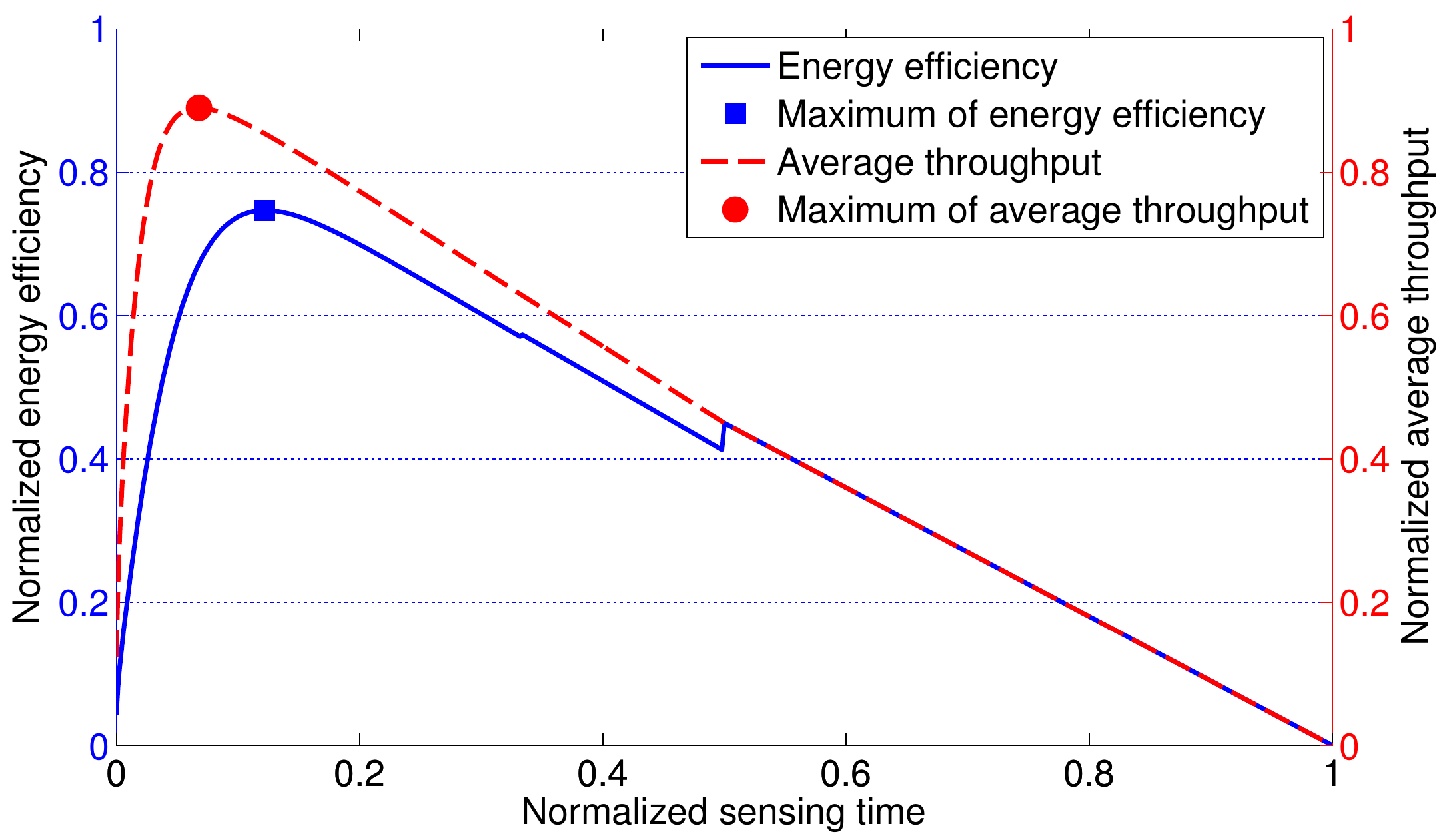}
		\label{subfig:SingleUserCase}
	}
	\subfigure[]{
		\includegraphics[width=8cm]{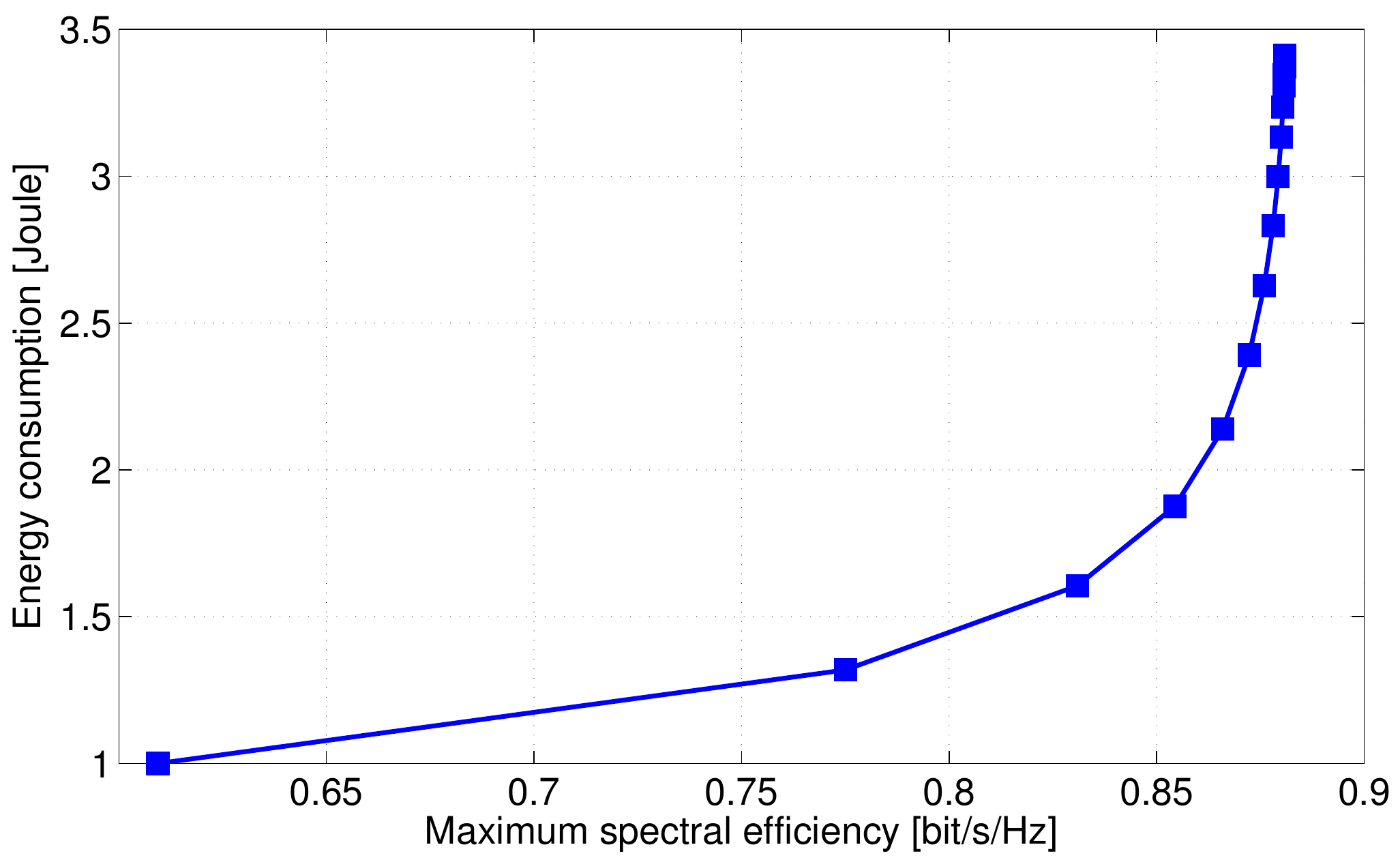}
		\label{subfig:EE_Thr}
	}	
	\caption{\subref{subfig:SingleUserCase} Energy efficiency versus sensing time in a multi-channel cognitive network with a single SU. The energy consumption, the denominator of the energy efficiency, is a non-continuous function, which causes the jump in the energy efficiency curve~\cite{ShokriIETC13}.
   \subref{subfig:EE_Thr} Throughput-energy tradeoff by increasing the maximum number of handoffs. Each dot corresponds to an increase of the number of handoffs. The energy consumption needs to be increased significantly to utilize all the available spectrum resources~\cite{ShokriIETC13}.}
	\label{fig:sensing-time}
\end{figure}

The interplay between sensing time, achievable throughput, and energy consumption for multi-channel system is evaluated in~\cite{ShokriIETC13}.
As shown in Fig.~\ref{subfig:SingleUserCase}, the energy efficiency first increases with the sensing time, due to a more accurate spectrum sensing, and reaches a maximum value. After this point, the energy efficiency falls, as the increased sensing performance cannot compensate for the increased energy consumption and for the decreased time available for transmission. The optimal sensing time for maximizing energy efficiency is higher than that for maximizing throughput, as it becomes more important to avoid false alarms and unnecessary additional sensing.

Secondary access without spectrum sensing (that is, zero sensing time) may improve the secondary throughput, if the primary system can tolerate some loss and the channel between the SU transmitter and PU receiver is expected to be weak, as shown in ~\cite{Fanous2014Access}. As this scheme at the same time introduces more packet loss in the secondary network with multiple uncoordinated SUs, its energy efficiency remains to be evaluated.


\subsection{Waiting or Handoff}
Once a PU returns to its channel, the SU may decide to wait until the channel becomes idle again, or invest some time and energy to start the spectrum handoff procedure and migrate to an idle channel.
As~\cite{WangVT12} suggests, the decision should be based on the throughput and delay requirements of the SU. Unless the secondary quality requirements are very strict, optimizing the probability of waiting instead of migrating can decrease the energy consumption of the SU by $20\%$. Clearly, the gain decreases as the throughput or delay requirements get strict, and the SU cannot afford to simply wait for the new transmission opportunity in the current channel.

Given that a waiting SU needs to discover that the channel becomes idle again,~\cite{zhang2013what} investigates how often the channel should be sensed. More frequent sensing allows the SU to start to transmit with lower discovery delay and thus to achieve higher throughput, at the cost of a higher sensing energy consumption. Sensing, however, does not need to be periodic. As it is shown in~\cite{zhang2013what}, the adaptation of the sensing interval, based on some knowledge on the PU busy time distribution can halve the discovery delay, and thus increase secondary throughput, under the same sensing energy budget as periodic sensing.

\subsection{Sensing Order}\label{subsec: sensing-order}
Under narrowband sensing, an SU sequentially senses the channels until an idle one is found. The order of the channels to be sensed affects the throughput and the energy consumption.
As a result of an improper sensing order, an SU may sense several channels to find a transmission opportunity, and thereby may suffer from more energy consumption and a shorter remaining transmission time.
Therefore, hybrid spectrum handoff strategies are considered in~\cite{PeiJSAC2011}, where the SU learns both the channel occupancy and the transmission channel quality statistics, and defines the sensing order accordingly. It is shown that optimizing only based on one of the above parameters can be highly sub-optimal, with a loss of energy efficiency up to $5-20\%$ for the considered scenarios.

Primary traffic shaping, for example, as a consequence of applied network coding, can increase the performance of learning the channel occupancy statistics, and can decrease the number of channels sensed until a transmission opportunity is found by as much as 50\%, leading to significantly improved energy efficiency, as shown in~\cite{fanous2014reliable}.


\subsection{Maximum Number of Handoffs}
The performance of narrowband sensing depends not only on the sensing time of a single channel and on the sensing order, but also on the number of channels that should be sensed before the SU stops searching for a while.
Clearly, allowing an SU to investigate more primary channels increases the chances of finding an empty one, leading to throughput enhancement. However, as we see in Fig.~\ref{subfig:EE_Thr}, the energy consumption cost of this increase can be tremendous, once the system is close to the throughput limit. For instance, to increase the throughput above $0.85$, only $3\%$ transmission rate enhancement is achieved by $81\%$ more energy consumption, which devastates the energy efficiency. This suggests that  the maximum number of possible handoffs need to be limited and the SU forced to wait, to improve the energy-throughput tradeoff.


\subsection{Sensing and Channel Access}
\begin{figure}[t]
	\centering
	\subfigure[]{
		\includegraphics[width=8cm]{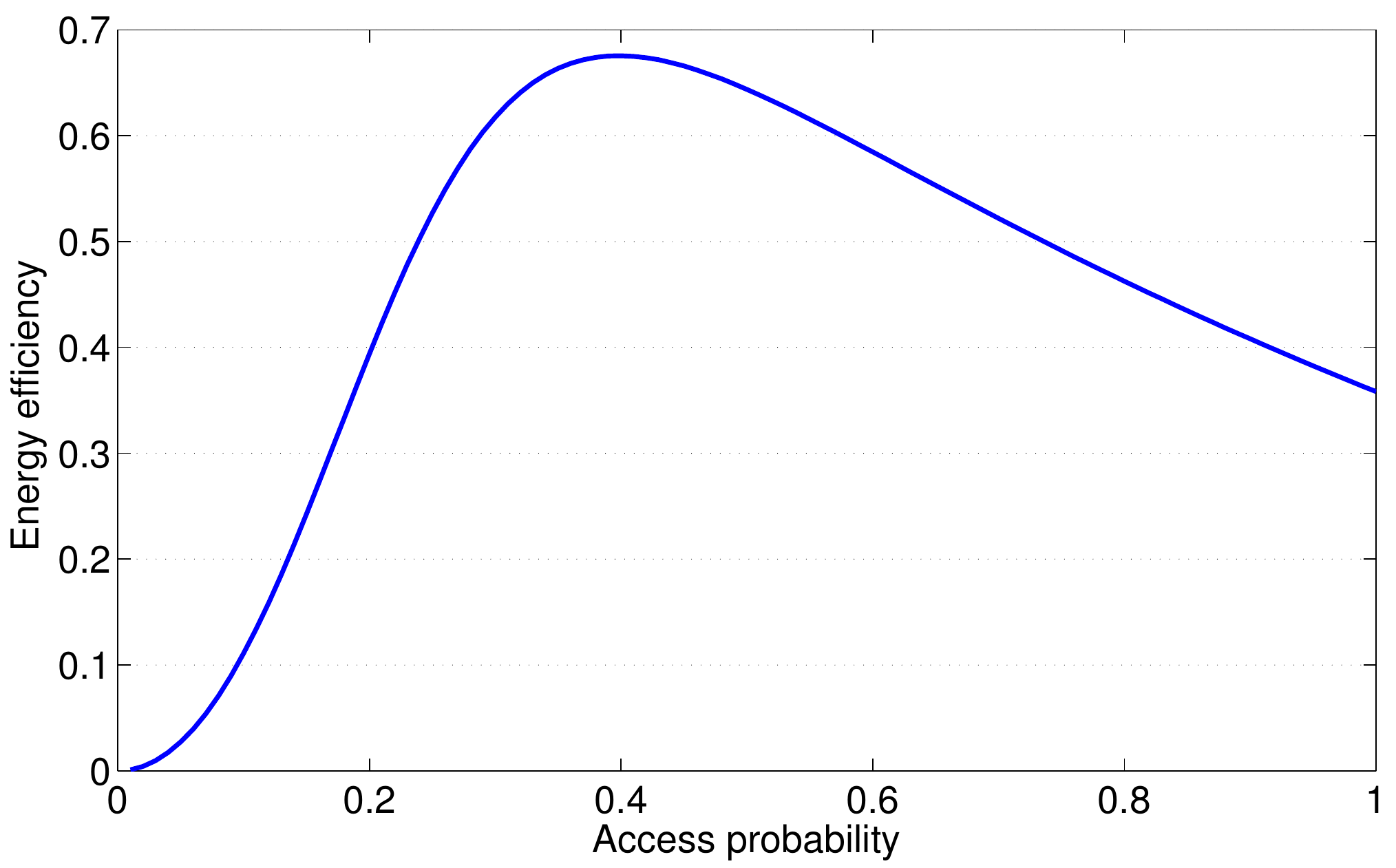}
		\label{subfig:EE_p}
	}
	\subfigure[]{
		\includegraphics[width=8cm]{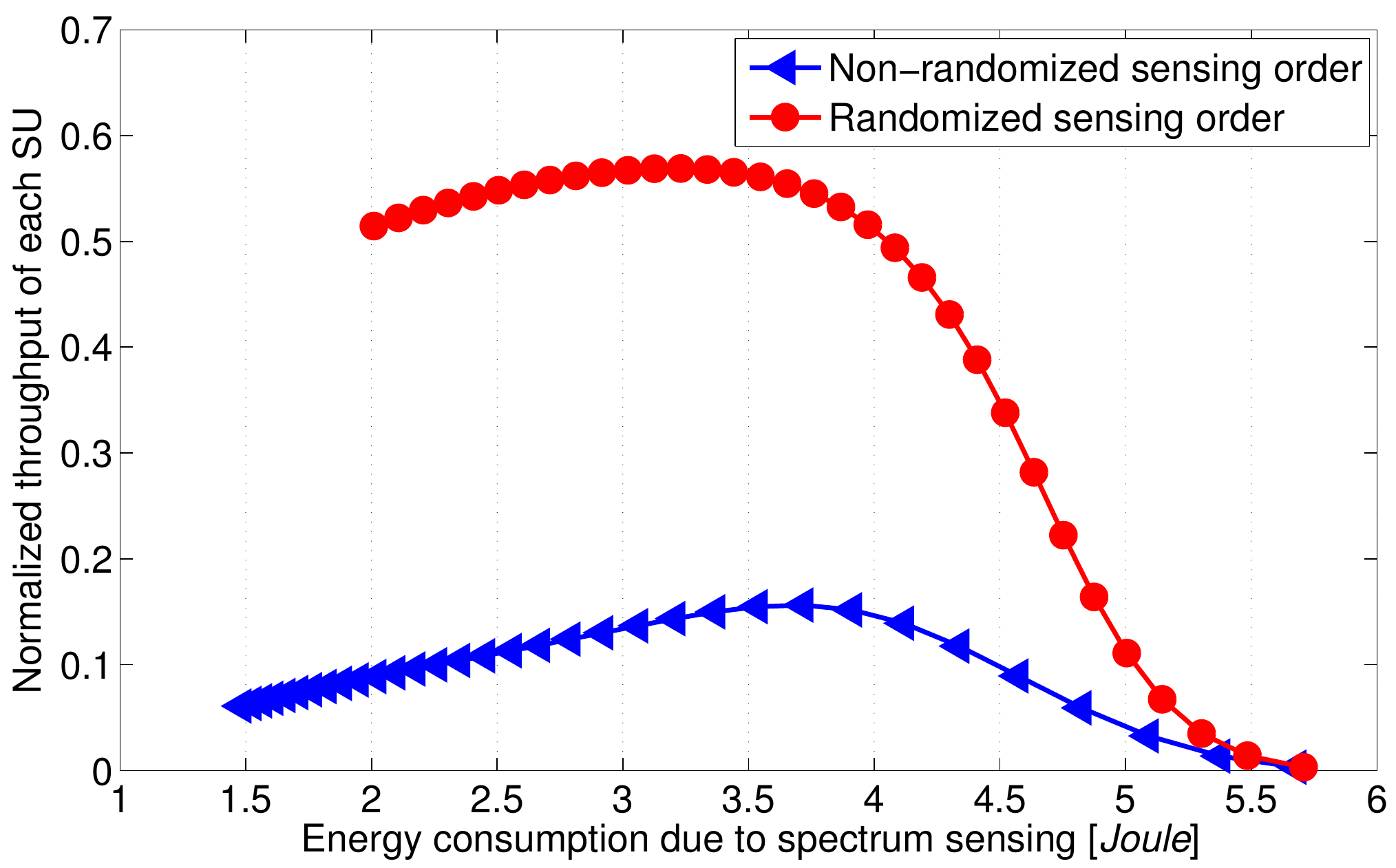}
		\label{subfig:thr_ho}
	}	
	\caption{\subref{subfig:EE_p} Energy efficiency as a function of secondary access probability. Optimizing the access probability can improve energy efficiency significantly.
   \subref{subfig:EE_Thr} Secondary throughput as a function of invested sensing energy. The joint access probability and randomization of the channel sensing order achieves significant throughput gain~\cite{Shokri2015Analysis}.}
	\label{fig:sensing-access}
\end{figure}

In a secondary network with several uncoordinated SUs, finding an idle channel does not guarantee successful transmission. Here, all SUs may sense the popular primary channels (like the ones with low load and good transmission quality), and then compete for accessing the same channel, while other channels might be idle. To solve this problem,~\cite{Shokri2015Analysis} suggests to couple sensing and channel access control, by introducing a randomized scheme, where the SUs sense and then access the channels with some access probability. As shown in Fig.~\ref{subfig:EE_p}, the access probability has a significant effect on the energy efficiency, due to the tradeoff between throughput enhancement at more intentions to access the channels and the consequent increase of the contention level. The optimum access probability depends on the size of the secondary and primary networks. Significant further gain can be achieved by randomizing the order of the channels to be sensed, as shown in Fig.~\ref{subfig:thr_ho}, as it avoids potential constant scheduling conflicts among SUs. Then, the joint optimization of sensing order and access strategy, based also on the channel occupancy statistics is a logical next step ~\cite{song2012prospect}. However, it requires precise SU synchronization and extensive signalling, which challenge the applicability in ad hoc setting.

%
%


\subsection{QoS Control and Cooperation}
If the primary throughput or delay requirements are not strict,
some controlled secondary interference can be accepted at the primary receivers. In this case secondary sensing and channel access control solutions can be parameterized such that the primary packet loss is kept at an acceptable level. As it is shown  in~\cite{Fanous2014Access}, such controlled interference can benefit the secondary network. Further gains can be achieved if interference and the consequent unsuccessful primary transmissions are compensated by cooperative relaying from the SUs. Therefore,~\cite{Costa2014Energy}
proposes cooperation on the network layer, which imposes only low signaling overhead, resulting in up to 50\% energy efficiency gain.

\section{Cooperative Spectrum Sensing}\label{sec: cooperative-SS}
In the case of strong primary signals, local sensing may be sufficient to ensure adequate sensing performance. However, the cooperation of several spectrum sensing devices, that is, SUs in the area, is needed if the primary signal is weak, or if the radio propagation environment is harsh. Under cooperation the spatial diversity among the SUs mitigates the effect of link impairments due to fading and shadowing, and the SUs, together, can more efficiently discover spectrum access opportunities. At the same time cooperative sensing introduces additional energy cost as local sensing results are reported to a central node, or shared among the SUs in the area.

The design factors discussed for local spectrum sensing can be also optimized in cooperative sensing scenarios, considering the wireless environment of the individual SUs. However, there are even additional open questions affecting the energy efficiency for cooperating users, as we discuss in the following subsections.

\subsection{Sensing Resource Allocation}\label{subsec_sensing_resource_allocation}
\begin{figure}[t]
	\centering
	\subfigure[]{
		\includegraphics[width=6.15cm]{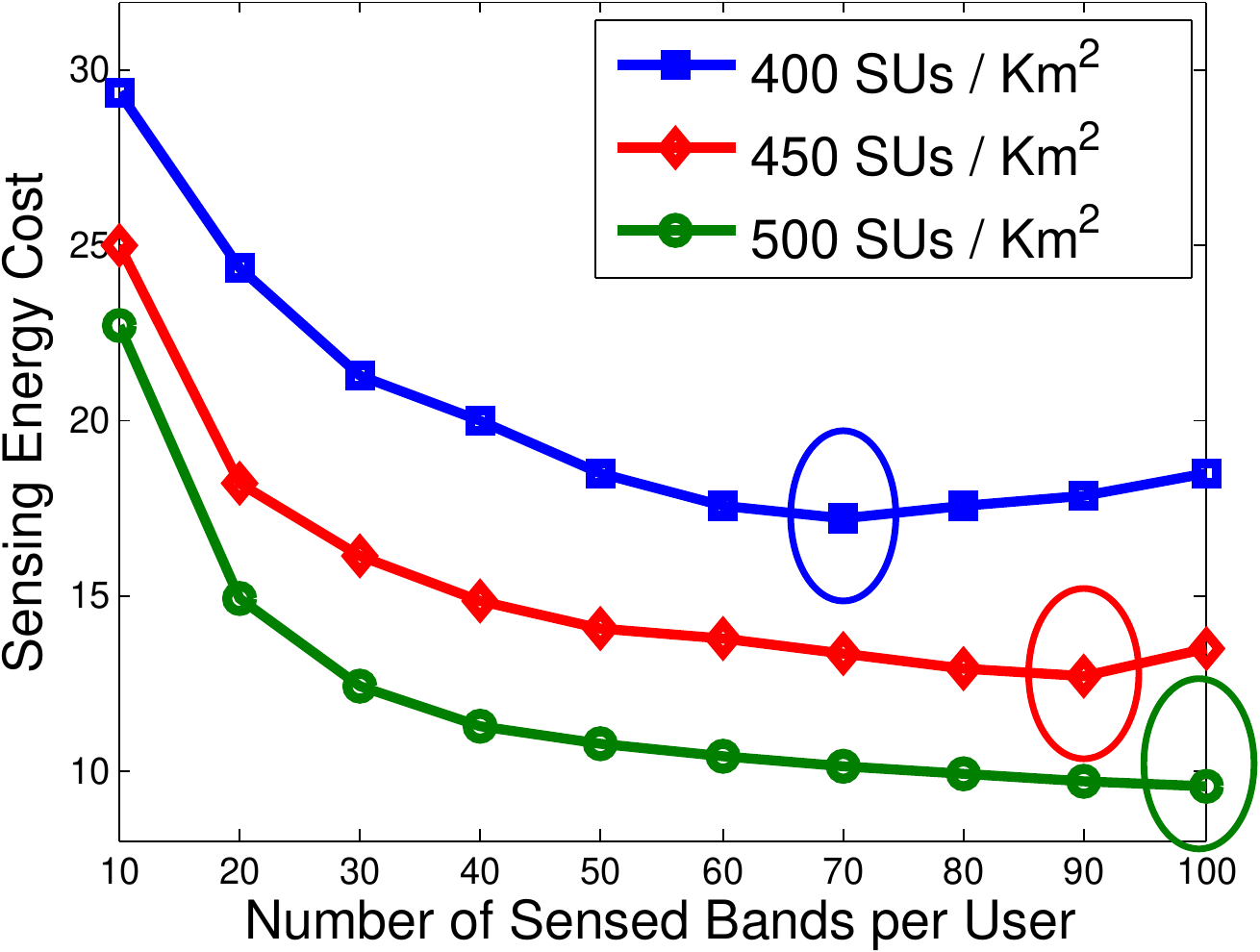}
		\label{subfig_energy_efficiency_vs_number_of_bands}
	}
	\subfigure[]{
		\includegraphics[width=6.15cm]{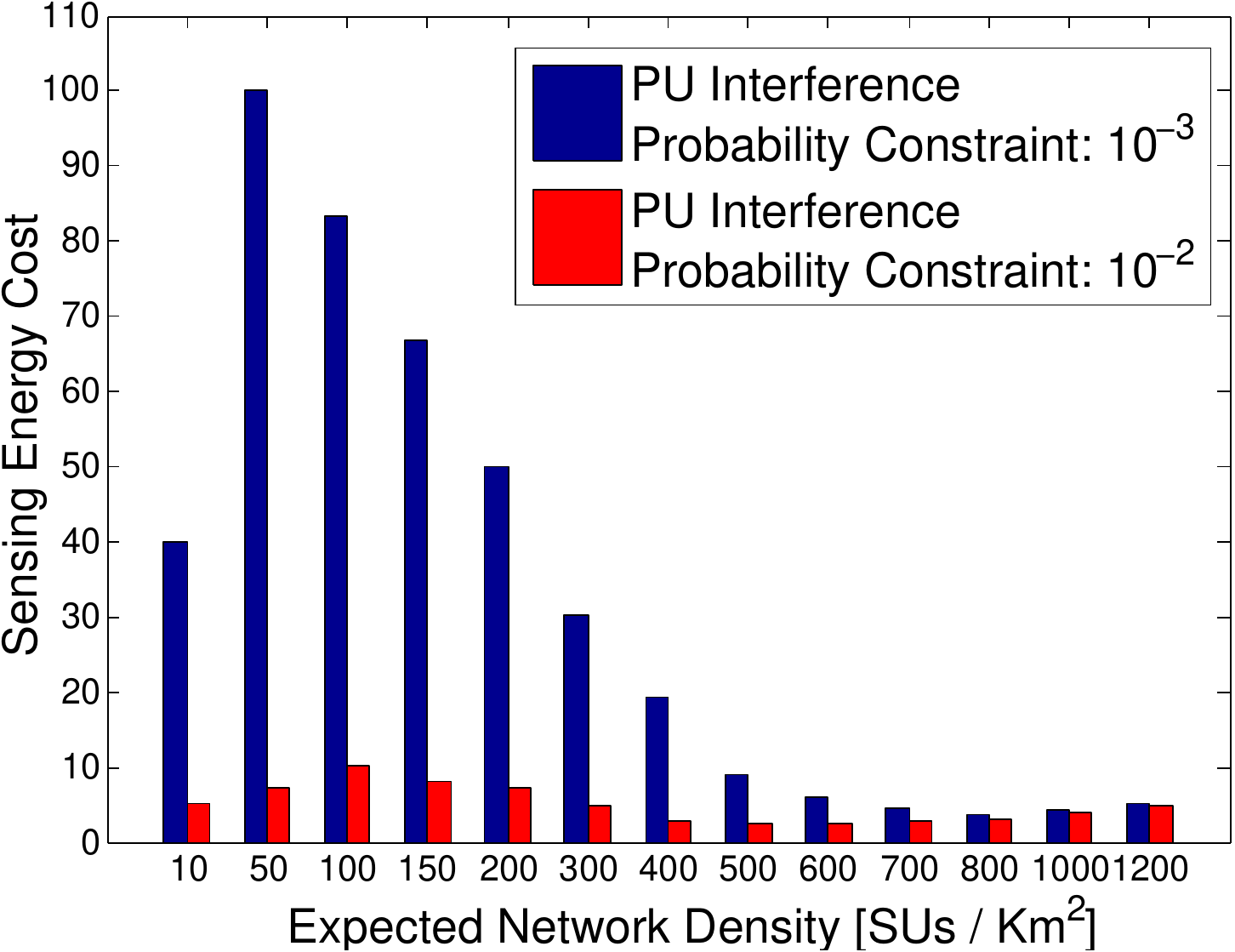}
		\label{subfig_energy_efficiency_vs_density}
	}	
    \caption{\subref{subfig_energy_efficiency_vs_number_of_bands} Average SU sensing energy cost per unit of data transmitted for each SU. The energy cost is minimized when the number of sensed channels is optimal. The optimum --indicated by a circle-- depends on the SU density. A small number of bands results in a
	lower achievable SU throughput, while exceeding the optimal spectrum space
	results in lower sensing efficiency per band, thus higher cost per achievable
	SU throughput.
	\subref{subfig_energy_efficiency_vs_density} The energy cost is lowest at
	optimal cognitive network density, above which the sensing performance improvement does not compensate for the increased demand for cognitive
	capacity~\cite{glaropoulos2014}.}
	\label{fig_glaropoulos}
\end{figure}

Under cooperative sensing, the sensing resource is not only the sensing time but also the set of SUs that cooperate to discover a spectrum access opportunity. Increasing the number of cooperating SUs may decrease the required contribution of each one of them, but may increase the overall energy consumption, or decrease the number of channels that can be sensed.
As the discovered spectrum access opportunities are used by the SUs themselves that are discovering spectrum opportunities, the SUs now need to decide how large part of the spectrum space, dedicated for secondary access, they want to utilize. On one side, they may want to increase the number of sensed channels, so that there are more transmission opportunities to share. On the other side, this either requires increased sensing efforts from each SU or results in a decreased per channel sensing accuracy under a constraint on the sensing cost of an SU. Consequently, there is an optimal value of the sensed channels that maximizes per SU throughput or minimizes the average SU energy cost  to achieve a transmission of a unit of data for each SU~\cite{glaropoulos2014}.  As it is shown in Fig.~\ref{subfig_energy_efficiency_vs_number_of_bands}, this optimal value depends on the network density. Moreover, as we see in Fig.~\ref{subfig_energy_efficiency_vs_density}, the energy cost, even if minimized, strongly depends on the primary network quality requirements as well as on the density of the secondary network. Networks with moderate density are worst off, where the cooperative sensing performance is moderate, but the gained access opportunities have to be shared by a relatively large set of nodes. Increased network density improves the energy efficiency significantly. Under very high densities the sensing energy cost increases again, as too many SUs need to share the low marginal sensing gain.

\vspace{-0.1cm}

\begin{figure*}[!t]
	\centering
	\subfigure[]{
		\includegraphics[width=5.75cm]{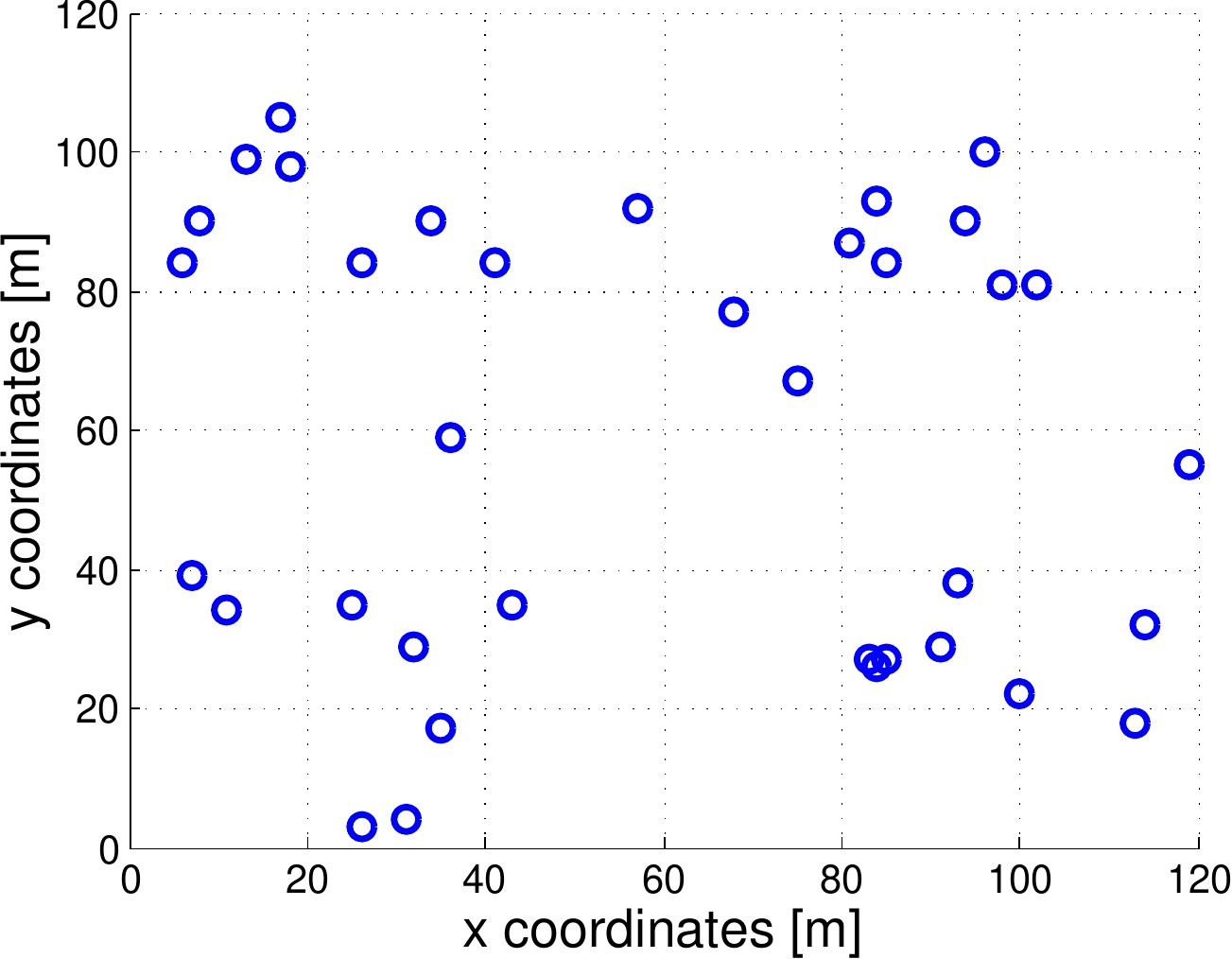}
		\label{subfig_topology_1}
	}
	\hspace{0.5cm}
	\subfigure[]{
		\includegraphics[width=5.75cm]{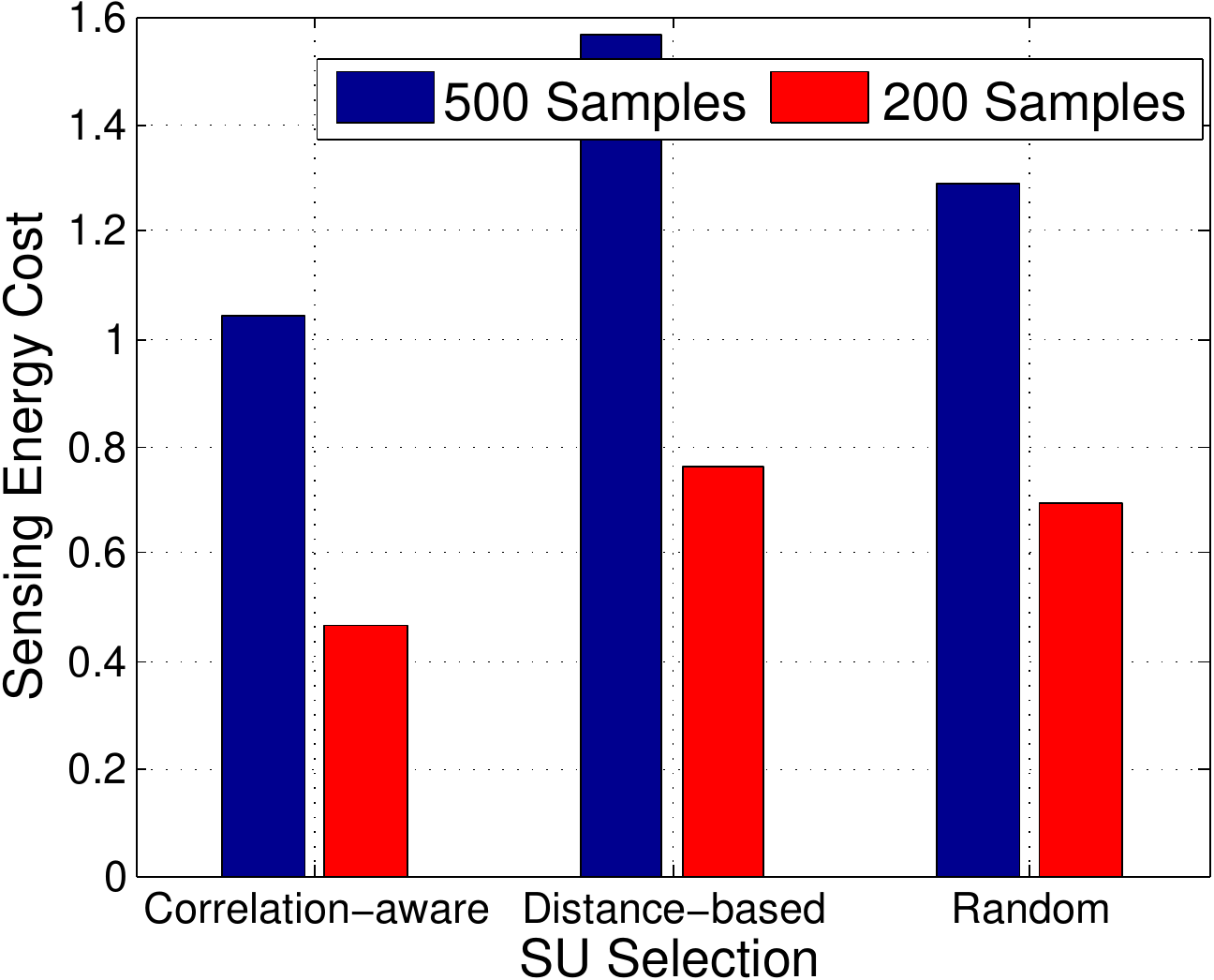}
		\label{subfig_energy_cost_vs_selection_1}
	}	
	\subfigure[]{
		\includegraphics[width=5.75cm]{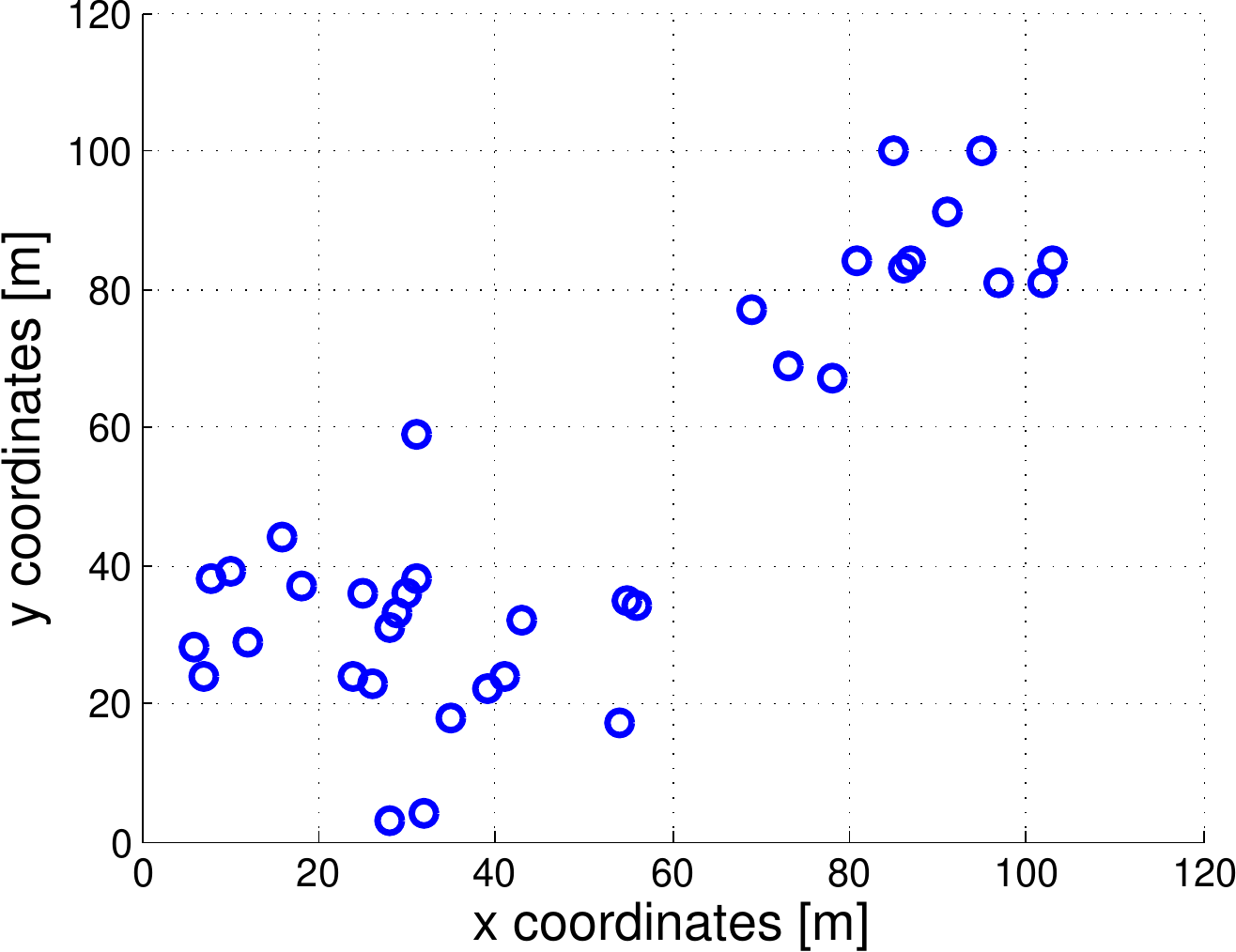}
		\label{subfig_topology_2}
	}
	\hspace{0.5cm}
	\subfigure[]{
		\includegraphics[width=5.75cm]{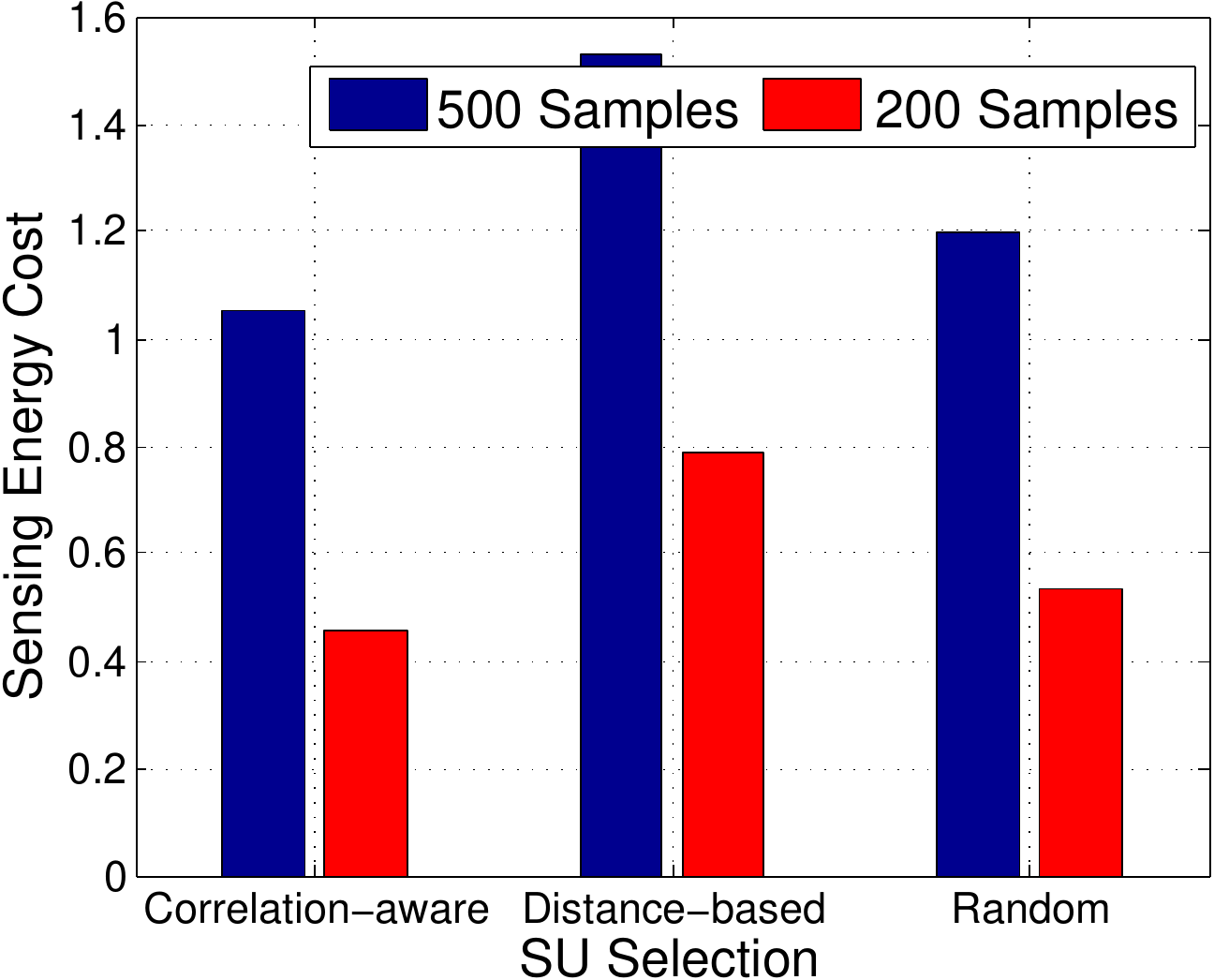}
		\label{subfig_energy_cost_vs_selection_2}
	}
	\caption{The energy cost per unit of SU throughput decreases when the correlation between SU channel measurements is taken into consideration in the iteration-based user selection algorithm. The improvement compared to random selection is smaller in \subref{subfig_energy_cost_vs_selection_2} since the nodes are located in disjoint geographical areas \subref{subfig_topology_2}. The higher sensing accuracy, as a result of the increase of the sensing time per channel (500 samples), does not compensate for the linear increase in the sensing energy overhead ~\cite{cacciapuoti2012}. }
	\label{fig_correlation_aware_selection}
\end{figure*}

\subsection{Sensing User Selection}\label{subsec_sensing_coordination_scheduling}
As the number of users participating in the cooperative sensing needs to be carefully selected, the remaining issue is which particular users should cooperate.

Accordingly,~\cite{saud2013} proposes an iterative solution to involve SUs in sensing, until the desired overall sensing performance in terms of misdetection and false alarm probabilities are met for all channels, with the objective of balancing the sensing energy consumption. Clearly, the gain of optimized SU selection increases together with the number of available SUs, and therefore is important in dense secondary networks.

Given that the main reason for cooperative sensing is to mitigate fading and shadowing,~\cite{cacciapuoti2012} suggests that users experiencing uncorrelated link
attenuation should be selected to cooperate. As shown in Fig.~\ref{fig_correlation_aware_selection}, the efficiency of this correlation-aware policy depends on the spatial distribution of the SUs. It can decrease the number of SUs required to sense the primary channel, and, consequently, the sensing energy cost by more than 50\%, without affecting the sensing accuracy.

\subsection{Sensing Report Forwarding}
\label{subsec_sensing_report_forwarding}

\begin{figure}[!t]
	\centering
	\subfigure[low SNR]{
		\includegraphics[width=6.0cm]{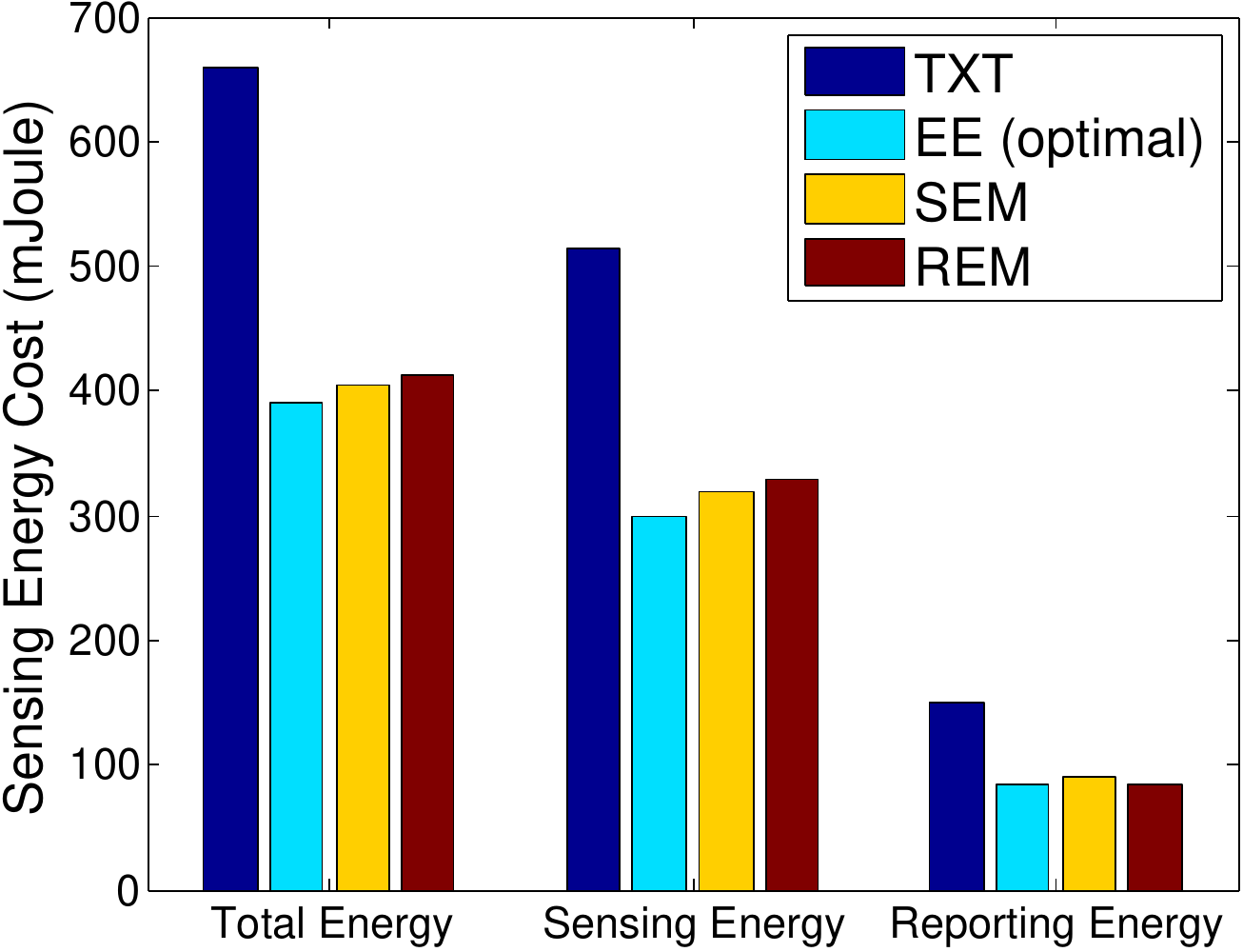}
		\label{subfig_energy_efficiency_vs_selection_low_snr}
	}
	\subfigure[high SNR]{
		\includegraphics[width=6.0cm]{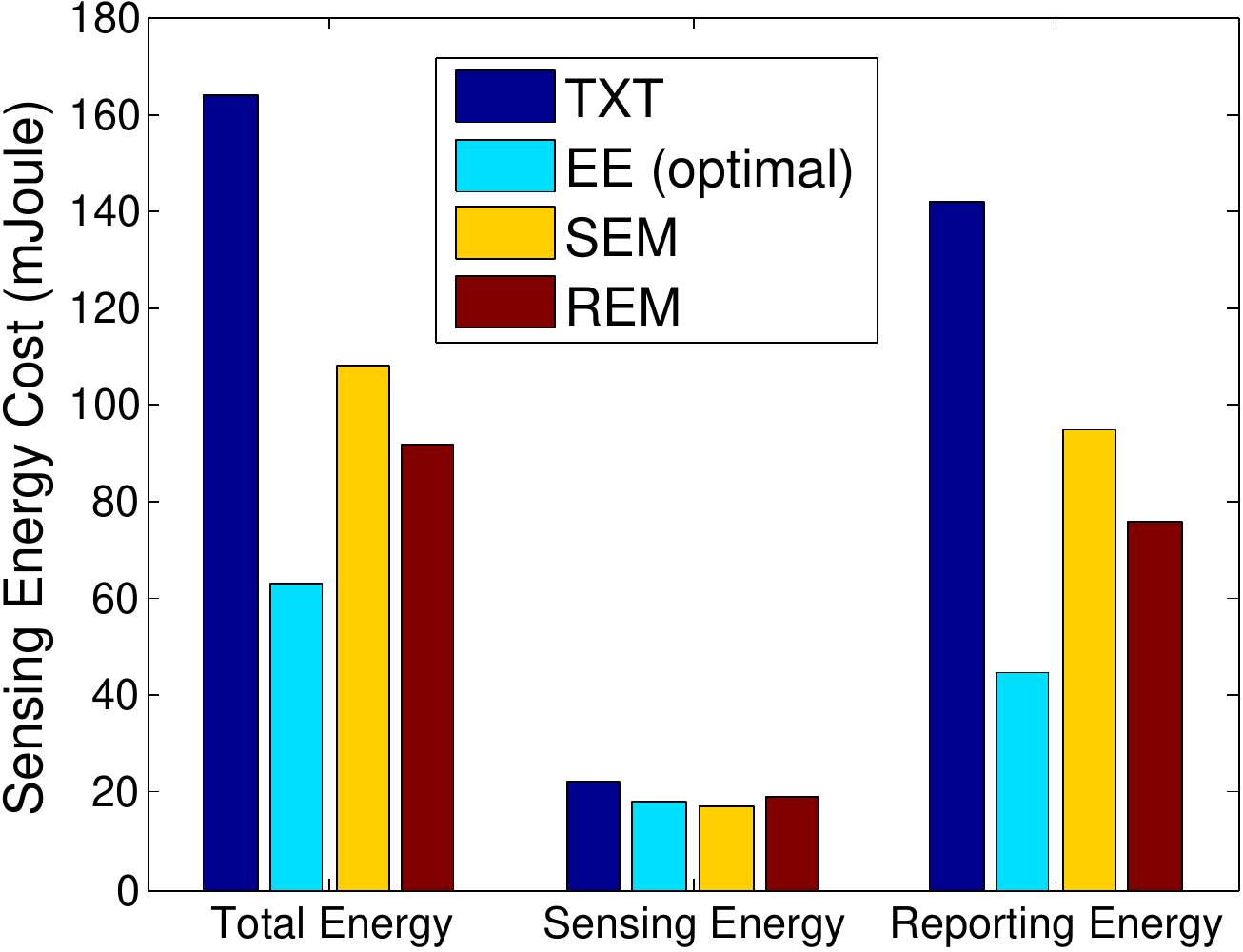}
		\label{subfig_energy_efficiency_vs_selection_high_snr}
	}	
	\caption{Cooperative sensing energy cost for the optimal and sub-optimal
	user selection schemes~\cite{salim2013}. In the low SNR regime
	\subref{subfig_energy_efficiency_vs_selection_low_snr} the sensing energy dominates the total energy consumption, while it drops significantly for high SNR
	\subref{subfig_energy_efficiency_vs_density}. The optimal (EE) SU selection scheme outperforms the heuristic solutions SEM and REM, which consider sensing energy cost and transmission costs respectively, and also TXT that minimizes the sensing time. The relative gain is more significant in high SNR regime, when sensing itself costs little energy.}
	\label{fig_energy_cost_vs_selection}
\end{figure}

Under cooperative sensing the local sensing results need to be reported, if centralized architecture is implemented, or shared among the neighboring SUs in distributed architectures. The reporting or sharing of the local sensing results may require a significant amount of energy and even time, particularly if high transmission power or multihop transmission is required. Therefore,~\cite{salim2013} compares different ways to select the cooperating SUs, considering only the sensing time (TXT), the local sensing performance (SEM), the sensing result transmission cost (REM), or all these (EE), with the objective to minimize the total required sensing energy cost for maintaining an overall sensing quality. As shown in Fig.~\ref{fig_energy_cost_vs_selection}, the gain
of joint optimization  is significant, if the sensing itself does not require much energy, that is, in the high SNR regime.

As reporting the sensing results may have significant cost,~\cite{maleki2011} suggests that the SUs, even if included in the cooperative sensing, should choose not to report the sensing result, if it might have little impact on the cooperative decision, while it would raise the overall reporting cost. The authors show that, if the primary channel utilization statistics are a-priori known to the SUs, then the individual SUs can have a good estimate on the validity of their sensing result. In this case, censored reporting can drastically reduce the total sensing energy overhead by up to 40\%, while the desired sensing performance is maintained.

%


In addition to introducing a significant overhead to the overall energy cost of collaborative sensing, the reporting of the individual sensing results may impose a threat to the cooperative sensing performance due to the inherent lack of reliability of the wireless links used for reporting. As ~\cite{chaudhari12} demonstrates, the quality of cooperative decision based on the individual decisions of the SUs (so called hard decision combining) can degrade by up to 60\% if the reporting links are unreliable. Instead, using cooperative decision based on quantized raw sensing results (that is, soft decision combining) the overall sensing performance can be maintained at a relatively high level. The granularity of the reported sensing results needs to be tuned carefully to tradeoff the energy cost and delay of reporting and the throughput gain due to correct spectrum decisions.

\section{Open Issues}
We have provided an overview of the most prominent mechanisms that aim to maximize the energy efficiency of spectrum sensing and handoff under local and under cooperative sensing. As of today the main focus of these various works is to characterize the achievable gains of these mechanism, under different networking scenarios, as summarized in Table \ref{tab: comparision}. However, to realize the predicted gains, several issues needs to be addressed by the research community.

\begin{itemize}
  \item Energy harvesting: emerging architectures with energy harvesting from interfering wireless signals change the general assumption of homogeneous energy resources at the nodes. To utilize energy harvesting, both local and coordinated sensing schemes need to be extended to consider the temporally and spatially varying harvested energy.
  \item Local sensing under dynamic traffic: most of the existing works considers SUs with saturated traffics and ideal wireless channel models, see~\cite{Shokri2015Analysis} and references therein. However, real network traffic is bursty, which makes it challenging to achieve the benefits of learning based system optimization,  due to the under-sampled or sparse network state information \cite{song2012prospect,fanous2014reliable,zhang2013what}.
  %
  \item Coexistence of SUs under local sensing: as SUs performing local sensing may belong to different networks, they may have no means or incentive to coordinate, and may have significantly different traffic demands and performance objectives. To take this heterogeneity into account, sensing and channel access optimization ~\cite{Shokri2015Analysis} needs to be extended with learning, fairness, and incentive mechanisms.
%
%
%
   \item Fair cooperative sensing: The optimization of the set of cooperating SUs, based on the sensing quality they can provide or the cost of communication ~\cite{cacciapuoti2012,salim2013,maleki2011}, may inherently lead to unfair allocation of sensing burdens in the cooperative systems. Future research is needed to evaluate whether this unfairness can be significant in fixed and in mobile environments, and how the performance of the proposed schemes changes if fairness is enforced, for example, considering a uniform sensing energy budget at the nodes, or contributions that are proportional to the needs of the individual SUs.
    \item Cooperative sensing incentives: Incentives are necessary to avoid free-riders, and, if possible, achieve a social optimum. Incentive schemes needs to be discussed considering short and long term objectives. On the short term, an SU may have incentive to cooperate if it has traffic to send, and needs free spectrum. Under dynamic traffic however, long term incentives need to be considered, to ensure that nodes cooperate, even if they do not have immediate gain.
\end{itemize}

\section{Conclusions}\label{sec: conclusions}
Improving the energy-throughput tradeoffs in spectrum sensing and access requires proper designs of the maximum number and the order of the primary channels sensed by an SU, the frequency of the spectrum sensing, and the selection of the per-channel sensing time to not let the energy resources be wasted for a marginally higher throughput. In multi-channel scenarios, the selection of the number and the order of the channels to be sensed becomes even more important.
In cooperative sensing scenarios, allocation of sensing tasks to SUs with relatively good individual sensing and uncorrelated channel conditions substantially reduces the energy consumption with negligible penalty in the network throughput. A careful reporting and combining of the individual sensing results, along with allocating sensing tasks to SUs with low-cost reporting links, increases the overall sensing and thereby energy efficiency.

\bibliographystyle{IEEEtran}
\bibliography{IEEEabrv,bibfile}
\end{document}